# Characterising the interplay between Galactic star formation and ionisation feedback with PRIMA

**Zavagno, A.**[a,b,*], **Russeil, D.**[a], **Suin, P.**[a], **Zhang, S.**[c], **Yadav, R.K.**[d], **Figueira, M.**[e,f], **Berthelot, L.**[g,a], **Arzoumanian, D.**[h,i,j], **Samal, M.R.**[k], **Rawat, V.**[k,l], **André, P.**[m], **Mattern, M.**[m], **Liu, H.-L.**[n], **Sadavoy, S. I.**[o], **Nozari, P.**[o], **Epinat, B.**[p]

[a]Aix Marseille Univ, CNRS, CNES, LAM Marseille, France
[b]Institut Universitaire de France, 1 rue Descartes, 75005 Paris, France
[c]Departamento de Astronomía, Universidad de Chile, Las Condes, 7591245 Santiago, Chile
[d]National Astronomical Research Institute of Thailand (NARIT), Sirindhorn AstroPark, 260 Moo 4, T. Donkaew, A. Maerim, Chiangmai 50180, Thailand
[e]Max-Planck-Institut für Radioastronomie, Auf dem Hügel 69, 53121, Bonn, Germany
[f]National Centre for Nuclear Research, Pasteura 7, 02-093, Warszawa, Poland
[g]Aix Marseille Univ, CNRS, LIS, Marseille, France
[h]National Astronomical Observatory of Japan, Osawa 2-21-1, Mitaka, Tokyo 181-8588, Japan
[i]Kyushu University, Department of Earth and Planetary Sciences, Faculty of Science, Nishi-ku, Fukuoda, Japan
[j]National Astronomical Observatory of Japan, Mitaka, Tokyo,
[k]Physical Research Laboratory, Navrangpura, Ahmedabad, Gujarat 380009, India
[l]Indian Institute of Technology Gandhinagar Palaj, Gandhinagar 382355, India
[m]Université Paris-Saclay, Université Paris Cité, CEA, CNRS, AIM, Laboratoire d'Astrophysique (AIM), Gif-sur-Yvette, France
[n]Yunnan University, School of Physics and Astronomy, Kunming, China
[o]Queen's University, Department of Physics, Engineering Physics and Astronomy, Kingston, Ontario, Canada
[p]Canada-France-Hawaii Telescope, Kamuela, Hawaii, United States

**Abstract.** Recent results from the James Webb Space Telescope show that nearby spiral galaxies are dominated by the presence of H I and H II bubbles that strongly shape their surrounding medium. These bubbles result from the feedback of high-mass stars at different stages of their life cycle. However, early (pre-supernova) feedback from high-mass stars is still poorly quantified. Recent results from numerical simulations suggest that the impact of high-mass star early feedback (photoionisation, wind) on star formation properties is complex, time-dependent and strongly depends on physical conditions, including the magnetic field properties. In our Galaxy, ionized (H II) regions observed in different evolution stages show a high diversity of star formation in their associated photo-dissociation regions (PDRs). However, the way in which the low- to high-density interstellar medium evolves to this situation remains elusive. Quantifying the impact of early feedback from high-mass stars on star formation properties and star formation laws (star formation rate, star formation efficiency versus gas surface density, $\Sigma_{\rm gas}$) will allow for a better understanding of the evolution of star formation laws in external galaxies, the laws that are key ingredients of galaxy evolution models. PRIMA, with its high sensitivity, large mapping efficiency and polarimetric capabilities, offers a unique opportunity to address the way radiative feedback and magnetic field control star formation in the Milky Way.

## 1 Introduction

Star formation occurs in dense and cold molecular clouds in galaxies. The *Herschel* space observatory 1 revealed the ubiquity of filaments 2, 3 that represent the structuring of interstellar matter in molecular clouds. Filaments host star formation above a column density threshold of about $7 \times 10^{21}$ cm$^{-2}$ corresponding to a background visual extinction of A$_V^{bg}$=7 mag 4. Note that this value is a not a sharp threshold and that lower values are reported in the literature [5–8]. The relation between feedback from high-mass stars, filaments and star formation has been established in several studies [9,10]. In particular, the high dynamics observed in high-mass star forming regions such as DR21 showed the existence of converging flows and their importance in the high-mass star formation process [11], the importance of global collapse of molecular cloud for high-mass star formation [12] and the central role of filamentary accretion flows in feeding star forming clusters [13].

Massive O and B stars ($M_* \geq 8$ $M_\odot$) have a profound impact on their surrounding medium all along their life through their radiation and wind. Recent JWST-MIRI images of nearby galaxies, such as that of the Phantom galaxy (NGC 628), show how important this impact is in shaping the surrounding molecular medium [14]. The effect of early radiative feedback on the distribution of molecular material has also been demonstrated in a larger sample of the nearby galaxies sampled by PHANGS ALMA and MUSE [15] on a spatial scale of 100 pc. However, in our Galaxy, despite dedicated observation programs such as FEEDBACK [16], little is known about the physics of this radiative feedback and its evolution as a function of time and the physical conditions of the medium in which star formation takes place. In particular, the effect that this feedback might have on star formation is highly debated. Is this effect constructive (favouring or accelerating the formation of new stars) or destructive (dissipating the gas and stopping further star formation)? Numerical simulations tend to conclude that this feedback is destructive [17,18] while observations indicate the opposite, even favoring the formation of a new generation of *high-mass* stars observed at the edges of ionized (H II) regions in the Galactic plane [19–22]. Note that this simplified view has to be refined because observational studies show that clouds are destroyed by stellar feedback while showing ongoing young high-mass star formation at the edges of the ionised regions [23] and that numerical simulations show the existence of star formation triggered by feedback before the cloud dispersal [24].

In many Galactic H II regions, the photodissociation region (PDR) that surrounds the H II region is observed as a dense layer of gas and dust where new stars form, including high-mass stars [20]. The supersonic expansion of the ionized gas in the surrounding medium creates a shock front that travels ahead of the ionisation front, allowing the formation of a dense layer of gas and dust between the two fronts. Under its own gravity this layer becomes unstable and fragments, forming dense clumps located in the PDR where a new generation of stars form. However, this situation strongly depends on local physical conditions and may not describe well the early stage development of the H II regions. This possible scenario was recently supported by a study of [25] that showed the dynamical impact of an expanding ionised region by observations of the [C II] line at 158$\mu$m with SOFIA and evidence for star formation triggered by the expansion of the ionised region. Another possibility is that the molecular could had already a clumpy structure and that self-gravity might be more important than feedback [17,26,27]. This question remains open.

Star formation occurs in dense clumps (of 0.3-3 pc) that fragment into cores (of 0.03-0.2 pc) [28]. The clumps and their associated cores are observed in filaments. The physical processes that start with the assembly of the interstellar medium into filamentary structures and extend to the formation of cores and subsequent star formation are still highly debated. In particular, the way low-density material ends up in dense structures is not well understood. Star formation observed in extreme environments, such as the Galactic center with SOFIA-FORCAST [29,30], or towards specific conditions such as those observed for Galactic cold cores [31] also reveal a complex process.

The way in which the star formation process is influenced by the physical conditions (level of tur-bulence, magnetic field) of its original medium also remains puzzling, even if recent work shows that these conditions clearly impact the properties of future star formation 21, 24, 32. In particular, the compression exerted on the surrounding medium by an expanding H II region increases the local density and might favor the formation of a new generation of stars 33 as observed in many Galactic 20 and extragalactic 34 star-forming regions.

The PRobe far-Infrared Mission for Astrophysics (PRIMA) is an infrared observatory (currently in Phase A) with a 1.8 m telescope actively cooled to 4.5 K. On board will be an infrared camera, PRIMAger, equipped with kinetic inductance detector (KID) arrays, covering the range 24 to 261 $\mu$m. PRIMAger provides two imaging modes: the hyperspectral mode will cover the 24-84 $\mu$m wavelength range with a spectral resolution of R$\simeq$10, while the polarimetric mode will provide polarimetric imaging in 4 broad bands from 80 to 261 $\mu$m (see L. Ciesla et al., this volume). The other instrument on board, the Far-Infrared Enhanced Survey Spectrometer Instrument, FIRESS (Bradford et al., this volume), covers the range 24 to 235 $\mu$m with four slit-fed grating spectrometer modules, providing a resolving power between 85 and 130, with a factor of 1,000 to 100,000 improvement in spatial-spectral mapping speed compared to *Herschel*. Compared to other infrared facilities, the high sensitivity and mapping efficiency of the PRIMA instruments offer new opportunities to study the feedback of high-mass stars on star formation in our Galaxy. For PRIMAger in particular, a Beam Steering Mirror (BSM) allows rapid, two-dimensional scanning of the instrument's field of view within the $42' \times 24'$ telescope field of view. Larger areas are imaged by combining this beam steering with the scanning motion of the entire observatory. While each detector images a different part of the PRIMAger field of view ($4' \times 4'$ for the individual arrays), the scanning strategy ensures that a common sky region can be imaged by multiple detectors (see Figure 1 in Ciesla et al., this volume). For PRIMAger, the combination of imaging and polarimetric modes in the 24-192 $\mu$m range is a great asset for the study of Galactic star formation. Key questions that PRIMA can address are

• How does compression by the incoming stellar radiation and the wind from the ionising star(s) forming the H II region affect the properties of filaments and clumps in the PDR around H II regions, at all evolutionary stages of the star formation process?

• How different are the properties of filaments and clumps in directly irradiated (facing the incoming radiation) and less irradiated zones of the PDR?

• How does the evolution of physical conditions (density, temperature, turbulence, magnetic field) affect the evolution of star formation in star-forming regions?

Section 2 presents some recent results on the feedback of high-mass stars, which raise new open questions that can be uniquely addressed with PRIMA. Section 3 outlines how PRIMA and its suite of instruments will be uniquely suited to characterise the feedback in star formation regions where different physical conditions and evolutionary stages are observed, and to follow the evolution of the star formation process from low to high density zones in our Galaxy. Section 4 summarises the results and concludes on the importance of the PRIMA mission for the study of Galactic star formation through the prism of early feedback from high-mass stars.

## 2 Recent feedback results and outstanding issues

*2.1 Molecular complexity in high-mass star-forming regions and the detection of faint sources*

Observations of different gas tracers in molecular clouds at different spatial resolutions reveal the complex structure of Galactic molecular clouds. Studies show that different evolutionary stages of star formation coexist at different spatial scales under the strong influence of H II regions 21, 35. Faint compact sources are observed in the far infrared with *Herschel*-SPIRE 36. These sources represent an early stage of star formation and their spectral energy distribution peaks in the 250–500 $\mu$m range. Their early evolutionary stage is confirmed by their detection in high density molecular gas tracers 37. The evolutionary stage of young stars detected in Hi-GAL, the *Herschel* survey of the Galactic plane 2 is based on their detection (or non-detection) at shorter wavelengths, in particular at *Herschel* PACS 38 70 $\mu$m 39. As explained in 39, this classification is first based on their detection at 70 $\mu$m, which indicates ongoing star formation and separates protostellar and starless sources. The sources are then classified using SED fitting and bolometric luminosity versus envelope mass, as first proposed in 40. PRIMA's higher sensitivity in the 70 $\mu$m range (2 orders of magnitude compared to *Herschel*-PACS) will allow the detection of faint, previously undetected sources. For all detected young stellar objects, a complete SED can be obtained and fitted with existing models 41 to derive their properties and evolutionary stages. A better census of the young stars/clumps population, in particular by accessing the faint sources, will significantly improve our estimate of the star formation laws (see section 2.2).

## 2.2 Star formation laws and the effect of feedback

In galaxies, the Kennicutt-Schmidt (KS) law relates the surface density of the star formation rate (SFR), $\Sigma_{SFR}$, to the local surface gas density, $\Sigma_{gas}$, by the relation $\Sigma_{SFR} \propto (\Sigma_{gas})^n$, where $n = 1.4 \pm 0.15$ 42, 43. In external galaxies, looking at spatial scales smaller than 100 pc, it has been proposed that the effect of feedback from high-mass stars modifies the surrounding medium and disperses molecular clouds on short timescales ($\sim$ 3 Myr), changing the KS relation 44. Recent high spatial resolution (0.007 pc) magnetohydrodynamic simulations of a high-mass ($10^4$ M$_\odot$) molecular cloud affected by high-mass stellar feedback and protostellar jets from young sources show that the presence of stellar feedback strongly influences the shape of the KS relation 24. In particular, this study shows that the effect of feedback on the KS relation is primarily determined by the way in which the feedback modifies the cloud structure. Figure 1 shows the evolution of the structure of a high-mass ($10^4$ M$_\odot$) cloud as a function of time, and how the gas is both dispersed and collected into denser material in a part of the cloud where star formation is occurring 24. Figure 1 shows the density projections along the z-axis for two snapshots of the numerical simulation presented therein, which includes both radiative feedback from an ionized region and protostellar jets associated with young stars forming in the cloud. The left part shows the evolution stage of the cloud at 2.3 Myr and the right part shows the evolution at 3.15 Myr. The white dots represent the positions of the sink particules. The left panel shows the formation of the central hub, into which three filaments converge. In the right panel we can see how the H II region, originating from the lower left filament, restructured the cloud after about 0.4 Myr. We show that the presence of stellar feedback strongly influences the shape of the KS relation and the star formation efficiency per free-fall time ($\epsilon_{ff}$). The effect of feedback on the relation is primarily determined by its influence on the cloud structure. Furthermore, the evolution of ($\epsilon_{ff}$) throughout the star formation event suggests that variations in this quantity may mask the effect of feedback in observational studies that do not take into account the evolutionary stage of the clouds. that observations tend to underestimate the total SFR. We also show that observations tend to underestimate the total SFR. Therefore, access to the faint population of young sources via high-sensitivity far-infrared observations is crucial for a better estimate of the SFR. Different structures of Galactic clouds have also been observed, where star-forming regions experiencing feedback show different density distributions and star formation law relations 45. The way in which the feedback of the high-mass star affects the surrounding medium depends strongly on the initial density distribution of the molecular material, which in turn is determined by the initial turbulence and magnetic field. This suggests that understanding the feedback requires a complete knowledge of the physical conditions, sampling a wide range of densities, spatial scales and evolutionary stages. This will only be possible with PRIMA.

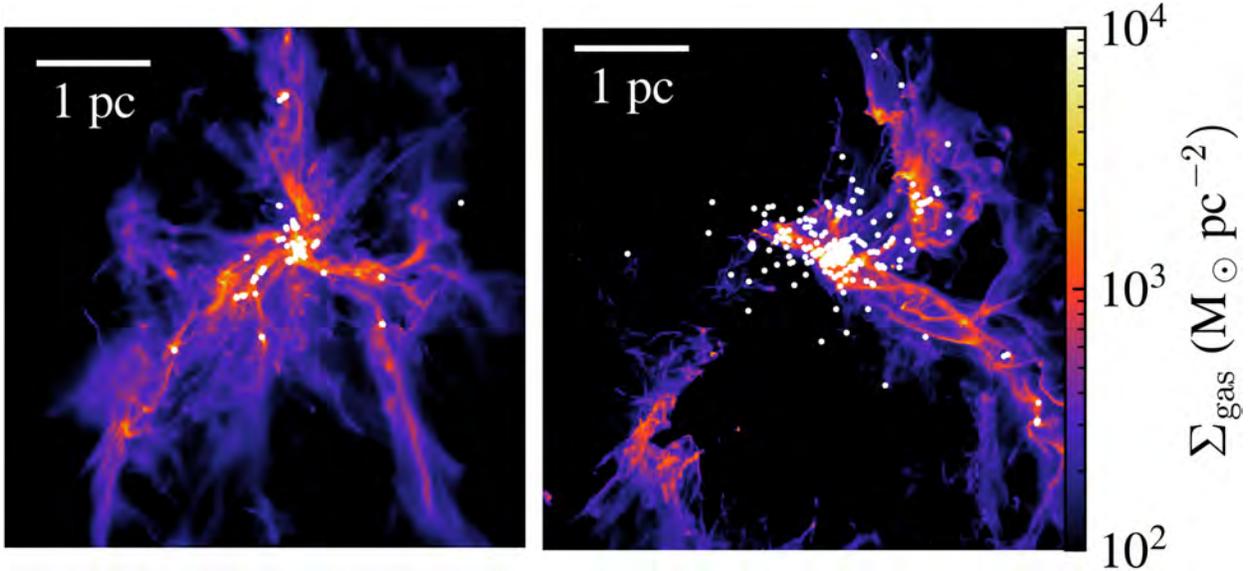

**Fig 1** Numerical simulation showing the density projections along the z-axis for two snapshots of the simulation including the effect of feedback from a 100 M$_\odot$ star and protostellar jets from young stars on a 10$^4$ M$_\odot$ molecular cloud at 2.3 Myr (left) and 3.15 Myr (right) (figure taken from 24 with permission of the authors).

### 2.3 Star formation is a highly dynamic process

Star formation is a highly dynamical process that occurs on multiple spatial scales. Several scenarios describe this multi-scale process to explain the formation of high-mass stars. These include the cloud-cloud collision, where the supersonic collision of molecular clouds produces a shock-compressed layer that leads to the formation of high-mass stars via gravitational instability, as proposed by 46. Another scenario based on numerical simulations proposes the formation of massive stellar clusters in converging galactic flows with photoionization 47. It is shown that more massive clusters form on shorter timescales for regions with strongly converging flows. The role of photoionisation feedback does not prevent massive cluster formation, but limits the masses of the clusters and has a stronger effect on lower density regions where there are no converging flows 47. Stars form in clusters where there are strong dynamical interactions. During the early stages of star formation, high-mass stars develop a H II region that will influence the future evolution of the cluster. However, depending on the initial density in the cluster, the evolution and expansion of the H II region may be delayed. In fact, high-mass protostellar sources with luminosities below 10$^5$ L$_\odot$ may take longer to develop a detectable ultra-compact ionised region, as observed by 48. Little is known about this process of very early feedback from massive stars in clusters.

Another open question is the effect of continuous material infall from the surrounding medium on a star-forming region. Hub filament systems are the places where the most massive stars ($M_* \geq 100\,M_\odot$) form [11–13,49]. The cluster of stars that forms at the centre of these hub systems benefits from an extra supply of molecular material from the surrounding filaments [50]. This alters the ongoing star formation there, and could also alter the central cluster's mass function as a function of time, as stars formed in the surrounding filaments associated with the hub fall into the cluster. Such a complicated situation is illustrated in Figure 2, which shows the 250 $\mu$m *Herschel*-SPIRE image of the Galactic hub filament system Monoceros R2 [50,51]. In addition to the H II region associated with the central cluster, other H II regions observed in the filaments associated with the central cluster also affect their immediate surroundings. Young stars at different evolutionary stages are also forming in these filaments, demonstrating the complex interplay between different feedbacks at different spatial scales and evolutionary stages in the same star-forming region [51,52].

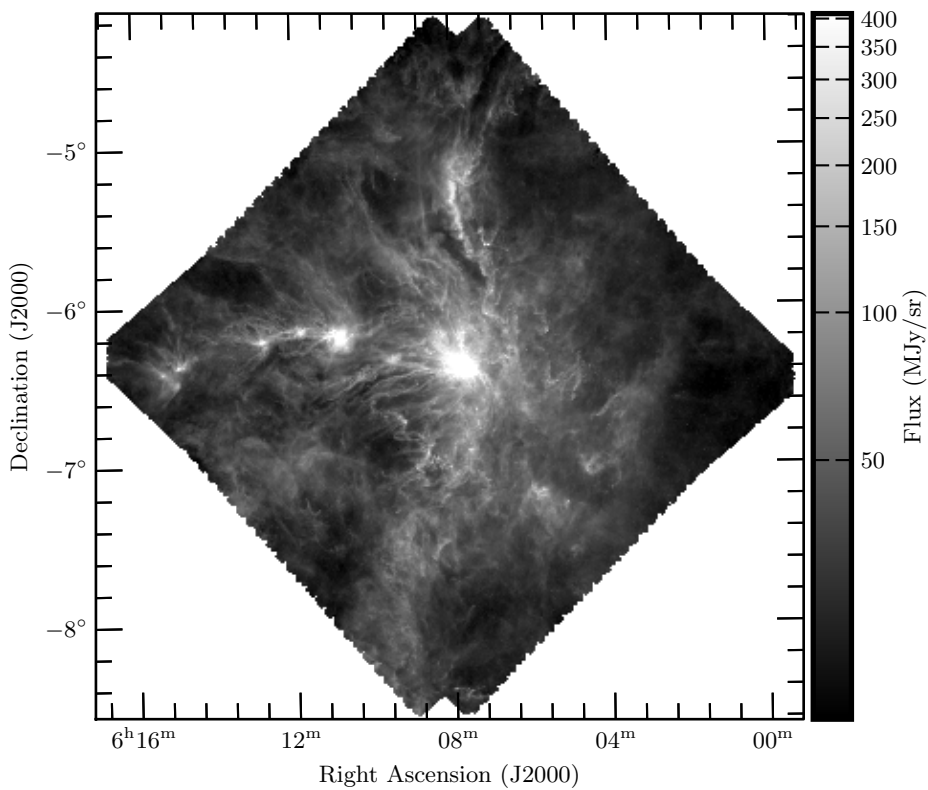

**Fig 2** The Galactic star-forming region Monoceros R2 observed with *Herschel*-SPIRE at 250 $\mu$m. The field size is 4.32°×4° (see also Figure 1 in [51] for a *Herschel*-PACS and SPIRE colour composite image of the region).

Thanks to its mapping efficiency and high sensitivity to detect faint sources over a wide wavelength range (25-300 $\mu$m), PRIMA will allow to sample the stellar population and the physical conditions over large fields/areas in complex star-forming regions.

## 2.4 Importance of the magnetic field

Many ground- and space-based facilities have been used to study the magnetic field in Galactic star-forming regions. These include early results from the Planck satellite [53], the SOFIA HAWC+ instrument to map the magnetic field structure of the OMC3 region, for example [54], the JCMT BISTRO survey [55], and dedicated large programs with ALMA such as the Magnetic Fields in Massive Star-forming Regions (MagMaR) project, which observed 30 high-mass star-forming regions at 1.2 mm polarisation (250 GHz) with ALMA [56] (and references therein). All these studies show the close relationship between the magnetic field and the star formation process, with a large diversity of observed situations depending on the evolutionary stage of the process, the density of the medium, the presence of radiative feedback [57] and the density tracers used [58]. The recent increase in the number of available measurements and dedicated studies further strengthens the interest in studying the magnetic field in Galactic star-forming regions [58–63]. In addition, numerical simulations show that the evolution of a massive star-forming clump is strongly influenced by the magnetic field and the radiative feedback [64].

The morphology of the magnetic field lines associated with H II regions has also been studied in detail, showing that the expanding H II regions push the gas and bend the magnetic field lines around them, locally modifying their initial orientation [65]. However, it is still difficult to measure the magnetic field in low density regions and to follow its evolution during the star formation process.

By being able to measure polarisation in the 80-260 $\mu$m range with high sensitivity, PRIMA will allow us to follow the evolution of the magnetic field from low to high density regions and to understand the role of feedback in this evolution (see section 3).

## 2.5 Triggered or not? Do massive stars trigger star formation?

Whether (or not) massive stars trigger star formation is still a matter of debate. The possible negative effects of feedback on star formation have been discussed in a number of papers. For example, [66], using velocity-resolved large-scale images in the fine structure line of ionised carbon ([C II] at 158 $\mu$m), show that the wind from the massive star $\theta^1$ Orionis C is responsible for the destruction of Orion's molecular core 1. This study provides observational evidence for the possible destructive effect of massive stars on the star formation activity in their birth molecular cloud. In addition, based on SOFIA FORCAST Giant H II survey data, [67, 68] show that the previously proposed sequential triggering star formation scenarios for W51A and M17 are not correct, because individual clumps show independent timelines of star formation histories in these two giant H II regions. However, a statistical study of 1360 Galactic H II regions, complemented by a number of dedicated individual studies of specific Galactic H II regions, has shown that there is an excess of very young star formation towards the PDR surrounding this large number of Galactic H II regions [20], raising the question of how radiative feedback affects star formation. Recent results indicate that high-mass stars begin to affect their surroundings very soon after their formation. ALMA and results from the ATOMS programme dataset [69, 70] show that cores and molecular material are observed around compact (<0.5 pc diameter) H II regions, probably due to their coeval presence with the high-mass star in the original cluster (see Figure 20 in [69] and Figure 14 in [70]). The difficulty of proving the causal link between an H II region and the young sources observed at its edges has always been a limiting factor in demonstrating the real impact of the ionised region in triggering the formation of a new generation of stars [71]. With PRIMA, the ability to sample a wide range of densities will allow all evolutionary stages to be sampled, from very compact to more extended H II regions. The high mapping efficiency of PRIMA will allow to study a large number of sources sampling different physical conditions and to follow the evolution of their associated environments. Access to the low-density material with PRIMA will be key to understanding the early stages of the star formation process.

## 2.6 The scope for deep learning

Because filaments drive star formation in galaxies, it is important to detect them in an unbiased way if we are to understand star formation. Classical extraction algorithms such as DisPerSE [72] or FilFinder [73] suffer from the need to define a threshold that makes this unbiased detection difficult over a wide range of filament densities and background contrasts. Using Hi-GAL $H_2$ column density images, [74] shows that the minimum column density that can be achieved for filament detection is about $10^{20}$ cm$^{-2}$. Note that this value depends on the location in the Galactic plane, reaching a value of $2\times10^{20}$ cm$^{-2}$ towards the Galactic centre (see Figure 2 in [74]). Recent studies using supervised deep learning on Hi-GAL $H_2$ column density images show that new filaments are being detected, in particular low density, low contrast filaments that were previously undetected.[75, 76]. These filaments could represent the first steps in the structuring of the low-density interstellar medium. The high sensitivity of PRIMA will allow these structures to be detected and their nature to be studied, shedding light on the first steps in the star formation process. The ability of PRIMA to combine the detection of filaments over a wide range of densities, including low densities, thanks to an improvement in sensitivity of 2 orders of magnitude compared to *Herschel*-PACS and SPIRE, together with the ability to observe the associated magnetic field, will open a new era in the study of Galactic filaments (see Molinari et al., this volume). This new data set available at the Galactic plane will allow the detection of faint structures and will help to confirm (or invalidate) the new detections obtained with supervised deep learning methods.

## 3 Need for PRIMA to quantify the effect of ionisation feedback on star formation in our Galaxy

PRIMAger, will provide observers with coverage of mid- to far-infrared wavelengths from about 25 to 264$\mu$m. PRIMAger will have two imaging modes: the hyperspectral mode will cover the wavelength range 25-80$\mu$m with a resolution of R=10, while the polarimetric mode will have four polarisation sensitive broadband filters from 80 to 264 $\mu$m. Thanks to its high sensitivity, efficient imaging power and polarimetric capability, PRIMA is particularly well suited to study the evolution of physical conditions in Galactic star-forming regions.

Active star-forming regions in the Galactic plane are bright in the infrared (from 30 to 2000 MJy/sr at 8 $\mu$m and from 20 to 1000 MJy/sr at 24 $\mu$m as observed with *Spitzer* and from 50 to 3500 MJy/sr at 250 $\mu$m as observed with *Herschel*-SPIRE. In addition, star-forming regions such as the NGC6334-NGC6357 complex are extended (typically $4° \times 2°$ in this case) and will require large maps to fully understand the star formation process there. PRIMAger offers a high mapping efficiency that will be invaluable for studying star formation in large Galactic complexes. PRIMAger will achieve a surface brightness sensitivity (total power I, background subtracted) in the hyperspectral imaging bands PHI1 and PHI2 of about 1 and 0.7 MJy/sr ($5\sigma$, 10h, 1 square degree), respectively (see Table 2 in Cielsa et al., this volume). This will allow the detection of the lowest mass/most embedded young stellar objects in the field for a new estimate of the star formation rate and efficiency in the clouds under (and away from) the influence of radiative feedback. PRIMA's combination of photometric imaging, polarimetry and far-infrared spectroscopy can address the important open questions presented in section 2. Of particular importance is the time evolution of the feedback effect of high-mass stars on low to high-density regions, and the role of the magnetic field in this evolution. The many galactic H II regions already observed in the optical and infrared will be observed with PRIMA to cover a wide range of geometries (compact, bipolar, extended) that are directly related to the physical conditions of the medium in which they form and to their stage of evolution. PRIMA will have the sensitivity and mapping capabilities to cover a large area (>1 square degree), allowing the necessary statistics to be achieved in terms of the Galactic regions covered to envisage quantifying the impact of radiative feedback on star formation. Combined with existing or upcoming surveys of the Galactic plane, a large plane coverage with PRIMA, combining high-sensitivity mapping and polarimetric measurements, would allow a significant leap forward in our understanding of the way Galactic star formation is regulated by the feedback from high-mass stars.

The high sensitivity of PRIMA's instruments will allow access to the faint and deeply embedded sources present in star-forming regions. This will allow a better estimate of the star formation laws, thanks to a more complete count of young stars/clumps, and will make it possible to study the dependence of their properties on the local physical conditions and the way in which they evolve with time 24.

The high sensitivity and efficient mapping of PRIMAger (Ciesla et al., this volume) will allow the evolution from the low to the high density medium to be linked during the star formation process. As can be seen in Figure 2, Galactic star-forming regions often exhibit, on a large spatial scale, a complex combination of young sources of different masses observed at different evolutionary stages, where the radiative feedback acts on different space and time scales.

The polarimetric capability of PRIMAger will be a strong asset for the study of Galactic star formation. Combined with the high mapping efficiency and sensitivity of PRIMA, this will allow the magnetic field to be studied over large areas and its role in the star formation process, with and without feedback, over a wide range of densities.

PRIMAger photometric and polarimetric data and FIRESS spectroscopic data, complemented by available data, will allow to
• Sample the young stellar/clump population in a region and determine their properties and evolutionary stages using dedicated spectral energy distribution models.
• Study the properties of the H II region using spectroscopy with FIRESS and short wavelength imaging with PRIMAger.
• To study the evolution of the magnetic field from low to high density regions thanks to the PRIMA polarimetric imager in the 80-260 $\mu$m range.

*3.1 Imaging*

The infrared domain (25-300 $\mu$m) covered by the PRIMAger Hyperspectral Imager (PHI) is particularly well suited to study the star formation process over a wide range of densities, including the low-density medium (N(H$_2$)=10$^{21}$ cm$^{-2}$). Star-forming regions located in (and above) the Galactic plane are bright in the infrared (from 30 to 2000 MJy/sr at 8 $\mu$m and from 20 to 1000 MJy/sr at 24 $\mu$m with *Spitzer* and from 50 to 8000 MJy/sr at 250 microns with *Herschel*-SPIRE). Some well-known nearby star-forming regions, such as the NGC6334-NGC6357 complex, are extended (typically 4°×2° in this case) and will require larger maps to allow a full understanding of the star formation process there. PRIMA's high-efficiency mapping provides such an opportunity and will be invaluable for studying star formation in large Galactic complexes.

Thanks to its high sensitivity and high mapping efficiency, PRIMAger will be able to map Galactic H II regions that sample different physical conditions (column density, temperature, turbulence level and magnetic field) in their surroundings. The surface brightness sensitivity that can be achieved in a reasonable time (total power I) in the hyperspectral imaging bands PHI1 and PHI2 of 4.5 and 2.5 MJy/sr ($5\sigma$, 10 hr, 1 sq deg) will allow the detection of the low mass/most embedded young stellar objects (with typical fluxes of 1 mJy at 30 $\mu$m and 2 mJy at 70$\mu$m) present in the field to obtain an unbiased census of the young stars/clumps population. The sensitivity of PRIMAger will be key to a new estimate of the SFR and star formation efficiency, and to discuss these quantities under (and without) the influence of ionisation feedback.

## 3.2 Polarimetry

The PRIMAger Polarimetric Instrument (PPI) [77] will allow us to understand how the evolution of magnetic field properties (orientation, intensity) affects the star formation process and how this evolution is related to the radiative feedback over a wide range of densities. This is particularly important in regions such as the hub-filament systems, where the feeding of fresh material by the filaments allows star formation in the central cluster to continue and be modified as a function of time.

The typical surface brightness sensitivity (polarised intensity P) between 0.65 MJy/sr in PPI1 and 0.25 MJy/sr in PPI4 ($5\sigma$, 10 hr, 1 sq deg, see Ciesla et al. this volume) will allow 5 to 10$\sigma$ detection in the low density regions (N(H$_2$)=$10^{21}$ cm$^{-2}$), where the polarisation fraction is expected to be around 1%, and a higher signal to noise ratio in regions where the polarisation fraction increases up to 8% [63].

## 3.3 PRIMA FIRESS

FIRESS (Bradford et al., this volume) will allow the physical conditions to be determined and followed from the spectral lines and continuum emission in the 24-235 $\mu$m range. This spectral range is ideal for observing the forbidden lines such as [O I] 63 $\mu$m, [O III] 88 $\mu$m [Si II] 35 $\mu$m and [C II] 158 $\mu$m which trace the PDR and characterise the physical conditions, including shocks and outflows, in star-forming regions using PDR models [78]. Both low and high resolution spectra can be obtained at key positions that sample the variety of conditions observed in star-forming regions, such as the ionised, neutral and molecular medium, with or without ongoing star formation. Looking at the evolution of physical conditions from low to high density regions affected (and unaffected) by ionisation feedback will be of particular interest to quantify the role of feedback. The FIRESS spectra will nicely complement the many *Herschel*-PACS and SPIRE FTS spectra obtained on Galactic H II regions and their associated PDR [79–81]. New spectroscopic results from the JWST will also be very important to get an overview of the physical processes at work in Galactic PDRs [82]. Regions observed with *Herschel* show bright lines and continuum (both in the PACS and SPIRE bands). Using the current estimate of $7 \times 10^{-19}$ W/m$^2$ in line emission ($5\sigma$, 1 hour sensitivity), a FIRESS spectrum covering the range 24 to 235 $\mu$m can be obtained in 2 hours.

## 4 Summary and Conclusions

Despite its clear importance, and as confirmed by recent JWST observations in nearby galaxies, the role of early radiative feedback on star formation in our Galaxy is still poorly quantified. PRIMA and its suite of instruments offer unprecedented advantages that will allow a significant leap forward in this field of research. PRIMA's high mapping efficiency combined with high sensitivity will allow it to cover large parts of the Galactic plane, providing the statistics and overview needed to quantify the feedback as a function of the parameters (physical conditions, evolution) that drive it. The polarimetric capability of PRIMAger will allow, for the first time, to study the role of the magnetic field in the star formation process over such a large and highly dynamic area. The unique opportunity provided by the combination of PRIMA instruments to study Galactic star formation over a wide range of densities will be key. Beyond this topic, and considering the proven importance of existing Galactic plane surveys and the many scientific topics they continue to serve long after their delivery, such an opportunity for a Galactic plane survey with PRIMA should really be considered.

*Disclosures*

The authors declare that there are no financial interests, commercial affiliations, or other potential conflicts of interest that could have influenced the objectivity of this research or the writing of this paper.

*Data availability*

No specific data are used for this publication.

*Acknowledgments*

AZ thanks the support of the Institut Universitaire de France (IUF). The authors thank the two referees for their comments that helped to improve the quality of the paper.

# List of Figures